\documentclass[twocolumn,english,aps,unsortedaddress,nofootinbib]{revtex4-2}

\usepackage{amsmath}
\usepackage{amssymb}
\usepackage[colorlinks]{hyperref}
\usepackage{epsfig}
\usepackage{graphicx}
\usepackage[T1]{fontenc}
\usepackage[utf8]{inputenc}
\usepackage{color}
\usepackage{booktabs,tabularx}
\usepackage{hyperref}
\usepackage[normalem]{ulem}
\usepackage{tikz}
\usetikzlibrary{positioning}

\begin{document} 

\title{Transfer-matrix approach to the Blume-Capel model on the triangular lattice}

\author{Dimitrios Mataragkas}

\author{Alexandros Vasilopoulos}

\author{Nikolaos G. Fytas}
\email{nikolaos.fytas@essex.ac.uk}

\affiliation{School of Mathematics, Statistics and Actuarial Science, University of Essex, Colchester CO4 3SQ, United Kingdom}

\author{Dong-Hee Kim}
\email{dongheekim@gist.ac.kr}

\affiliation{Department of Physics and Photon Science, Gwangju Institute of Science and Technology, Gwangju 61005, Korea}

\date{\today}

\begin{abstract}
We investigate the spin-$1$ Blume-Capel model on an infinite strip of the triangular lattice using the transfer-matrix method combined with a sparse-matrix factorization technique. Through finite-size scaling analysis of numerically exact spectra for strip widths up to $L = 19$, we accurately locate the tricritical point improving upon recent Monte Carlo estimates. In the first-order regime, we observe exponential scaling of the spectral gap, reflecting the linear growth of interfacial tension as the temperature decreases below the tricritical point. Finally, we validate our tricritical point estimate through precise agreement with conformal field theory predictions for the tricritical Ising universality class. Our results underscore the continued utility of the transfer-matrix approach for studying phase transitions in complex lattice models.
\end{abstract}

\maketitle

\section{Introduction}
\label{sec:intro}

The Blume-Capel model, originally introduced in Refs.~\cite{blume,capel}, describes a spin-$1$ Ising system with a single-ion uniaxial crystal-field anisotropy. Despite its apparent simplicity, the model has been extensively studied in both statistical and condensed matter physics due to the richness of its phase diagram and its broad theoretical appeal. Its relevance is further highlighted by its applicability to a variety of physical systems, including multi-component fluids, ternary alloys, and $^3$He–$^4$He mixtures~\cite{lawrie}. More recently, the Blume-Capel model has found renewed interest in diverse contexts such as the study of ferrimagnetism~\cite{selke-10}, wetting phenomena and interfacial adsorption~\cite{fytas13}, as well as investigations into the scaling behavior of the zeros of the partition function~\cite{leila} and the energy probability distribution~\cite{macedo24}.

\begin{figure}
    \centering
    \includegraphics[width=1.0\linewidth]{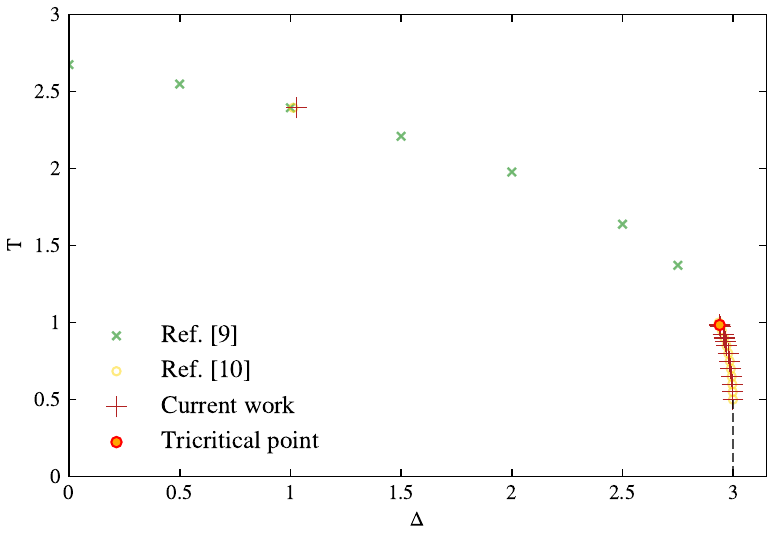}
    \caption{Phase diagram of the two-dimensional spin-$1$ Blume–Capel model on the triangular lattice in the $(\Delta, T)$ plane. Transition points obtained in the present work are shown alongside data from previous numerical studies using Wang–Landau and multicanonical simulations~\cite{fytas_BC,mataragkas2025}. An excellent overall agreement is observed between the different estimates across both segments of the phase boundary. The tricritical point $(\Delta_{\rm t}, T_{\rm t}) = [2.938568(1), 0.9847240(2)]$, as determined in this study, is highlighted by a red circle. The dashed line serves as a visual guide, indicating the first-order transition regime. For brevity, we set $J = 1$ and $J/k_\mathrm{B} = 1$ as the unit of energy and temperature hereafter.}
    \label{fig:pd}
\end{figure}

The zero-field model is described by the Hamiltonian
\begin{equation}
\label{eq:Hamiltonian}
    \mathcal{H} = -J \sum_{\langle i, j \rangle} s_i s_j + \Delta \sum_i s_i^2,
\end{equation}
where the spin variables $s_{i}$ take on the values $\{-1, 0, +1\}$, $\langle ij\rangle$ indicates summation over nearest neighbors, and $J > 0$ is the ferromagnetic exchange interaction. The parameter $\Delta$, known as the crystal-field coupling, controls the density of vacancies ($s_{i}=0$). For $\Delta\rightarrow -\infty$, vacancies are suppressed and the model maps onto the simple Ising ferromagnet. 
The phase diagram of the Blume-Capel model in the crystal field–temperature ($\Delta, T$) plane features a boundary separating the ferromagnetic and paramagnetic phases. At high temperatures and low crystal fields, this boundary corresponds to a continuous phase transition in the Ising universality class. In contrast, at low temperatures and high crystal fields, the transition becomes first-order~\cite{blume,capel}. As a result, the model provides a canonical example of a system exhibiting a tricritical point $(\Delta_{\rm t}, T_{\rm t})$, where the line of continuous transitions meets the first-order segment~\cite{lawrie}. For a recent high-precision determination of the phase diagram of the two-dimensional square-lattice Blume–Capel model, we refer the reader to Ref.~\cite{zierenberg17}. For the triangular-lattice version studied in the present work, an overview of the phase diagram--based both on our results and the computations reported in Refs.~\cite{fytas_BC,mataragkas2025}--is presented in Fig.~\ref{fig:pd}. A detailed compilation of transition points in the first-order regime, extending up to the tricritical point, is provided in Table~\ref{tab:pd}.

Since it was first proposed, the Blume-Capel model~\eqref{eq:Hamiltonian} has been investigated through mean-field theory, perturbative expansions, and numerical simulations on a variety of lattices, primarily in two and three dimensions; see, e.g., Refs.~\cite{fytas_BC,fytas2012,fytas2013,zierenberg2015}. The vast majority of studies have focused on the two-dimensional square lattice, employing a broad range of methods. These include real-space renormalization~\cite{berker1976rg}, Monte Carlo simulations and Monte Carlo renormalization-group approaches~\cite{landau1972,kaufman1981,selke1983,selke1984,landau1986,xavier1998,deng2005,silva2006,hurt2007,malakis1,malakis2,kwak2015}, $\epsilon$-expansion techniques~\cite{stephen1973,chang1974,tuthill1975,wegner1975}, high- and low-temperature series expansions~\cite{fox1973,camp1975,burkhardt1976}, and transfer-matrix calculations~\cite{beale1986,kim17,xavier1998,qian05,blote19}.

In the present paper, we investigate the {tricritical 
properties of the Blume-Capel model on the triangular lattice. For this particular lattice geometry, Mahan and Girvin~\cite{mahan} were the first to apply position-space renormalization-group methods, estimating the tricritical point at $(\Delta_{\rm t}, T_{\rm t}) = (2.686, 1.493)$. Many years later, Du \emph{et al.}~\cite{du} performed a more sophisticated analytical treatment using an extended Bethe-Peierls approximation, obtaining the estimate $(\Delta_{\rm t}, T_{\rm t}) = (2.841, 1.403)$. A further refinement, based on Monte Carlo simulations and an approximate scaling analysis of the specific heat, yielded $(\Delta_{\rm t}, T_{\rm t}) = [2.925(8), 1.025(10)]$~\cite{fytas_BC}. Most recently, Ref.~\cite{mataragkas2025} employed Wang-Landau simulations to determine the joint density of states, in combination with  histogram reweighting and field-mixing techniques based on the Metropolis algorithm. This approach led to a high-precision determination of the first-order transition line and the tricritical point, now located at $(\Delta_{\rm t}, T_{\rm t}) = [2.9388(2), 0.9850(8)]$.

Methodologically, we employ the transfer-matrix method with large strip widths. As will become evident below, this approach enables direct estimation of the phase diagram along multiple directions and, to the best of our knowledge, has not previously been applied to the triangular-lattice Blume-Capel ferromagnet. It proves particularly valuable in exploring the first-order transition regime, leveraging the deterministic and numerically stable nature of the transfer-matrix method to compute the free energy of strip systems without the complications arising from energy barriers--issues that often challenge even advanced Monte Carlo methods. In addition to its effectiveness in the first-order regime, the transfer-matrix approach enables precise determination of the tricritical point and supports detailed comparisons with exact results from conformal field theory. However, a well-known limitation of the transfer-matrix method lies in its poor scalability: finding the largest eigenvalues becomes computationally demanding as the strip width increases. This limitation underlines the importance of efficient matrix algorithms and significant computational resources. Given the recent advances in Krylov subspace algorithms~\cite{kim17} and the availability of much greater computational power, it is timely to revisit and update transfer-matrix calculations. In particular, this study reassesses the practical applicability of the transfer-matrix method around the area of the tricritical point through direct comparison with recent high-accuracy Monte Carlo results~\cite{mataragkas2025}.

In the transfer-matrix analysis, phase transitions are signaled by the asymptotic degeneracy of the largest eigenvalues of the transfer matrix. Within this framework, we closely examine the behavior of both the correlation length ($\xi_{L}$) and the persistence length ($\tilde{\xi}_{L}$). Using the persistence length, we construct a detailed phase coexistence curve and analyze thermodynamic properties in the first-order regime. Notably, we observe that the interfacial tension along the coexistence line increases linearly with the temperature as the system moves deeper into the first-order region, in agreement with previous results for both the square- and triangular-lattice Blume–Capel models~\cite{kim17,mataragkas2025}. We accurately determine the location of the tricritical point, finding excellent agreement with the most advanced Monte Carlo results to date~\cite{mataragkas2025}. Furthermore, we perform a stringent test of conformal field theory predictions by extracting the central charge and scaling dimensions from the transfer-matrix spectrum at our estimated tricritical point. These results strongly support the identification of the tricritical point with the Ising tricritical universality class and align closely with theoretical expectations~\cite{henkel_book}.

The remainder of this paper is organized as follows. In Sec.~\ref{sec:TM}, we outline our implementation of the transfer-matrix method for the Blume–Capel model on an infinite strip of the triangular lattice. Section~\ref{sec:results} presents a detailed finite-size scaling analysis of the persistence and correlation lengths, including a characterization of the first-order regime and a highly accurate
determination of the tricritical point, corroborated by precise verification
against conformal field theory predictions. Finally, Sec.~\ref{sec:conclusions} summarizes our main findings and discusses potential avenues for future research.

\section{Transfer-Matrix Method}
\label{sec:TM}

In this section, we present our implementation of the transfer-matrix method for the Blume-Capel model on an infinite strip of the triangular lattice. To construct a symmetric transfer matrix, we adopt the three-row geometry of double-layered triangles, as illustrated in Fig.~\ref{fig:tm-method}(a), and impose periodic boundary conditions along the transverse direction. For a strip of size $L \times M$, the partition function is expressed as $Z = \mathrm{Tr}\, \mathbf{T}^M$, where the transfer matrix $\mathbf{T}$ has dimensions $3^L \times 3^L$ in the spin-$1$ case. This exponential scaling quickly renders full diagonalization computationally prohibitive. However, since only the largest eigenvalues of $\mathbf{T}$ are needed in the limit $M \to \infty$, we avoid full diagonalization by employing sparse matrix factorization combined with the thick-restart Lanczos algorithm~\cite{TRLan}. This approach allows us to reach strip widths up to $L = 19$ within our available computational resources. 

Our sparse matrix factorization builds on the scheme introduced by Bl\"{o}te, Wu, and Wu for the Ising model on the honeycomb lattice~\cite{blote1990}, which we reformulate here for
the triangular-lattice Blume-Capel model. In the chosen three-row geometry of double-layered triangles, the transfer matrix $\mathbf{T}$ can be factorized as
\begin{equation} \label{eq:tmatrix}
    \mathbf{T} = \mathbf{R} \left( \mathbf{V}^{1/2}
    \mathbf{T}_{L} \mathbf{T}_{L-1} \cdots \mathbf{T}_2 \mathbf{T}_1
    \mathbf{V}^{1/2} \right)^2 \,,
\end{equation}
where $\mathbf{V}$ is a diagonal matrix that encodes intra-row interactions and local crystal-field contributions, and each $\mathbf{T}_k$ is a sparse matrix responsible for transferring the spin state at the k\textsuperscript{th} site. The full transfer through two stacked triangle layers is achieved by applying this sequence twice. Because the successive local transfers proceed along the diagonal direction, they introduce a one-site shift in the top row; the final operation, $\mathbf{R}$, performs a cyclic rotation of the spin configuration, restoring symmetry between the top and bottom rows, as depicted in Fig.~\ref{fig:tm-method}(a).

To define the matrix elements, we represent the spin configuration of a row of length $L$ using the basis ket $|\mathbf{s} \rangle \equiv |s_1 s_2 s_3 \cdots s_L \rangle$, where $s_j \in \{-1, 0, 1\}$ denotes the spin at the j\textsuperscript{th} site, counted from the left. The index $\mathbf{s}$ corresponds to a ternary encoding, given by $\mathbf{s} = \sum_{j=1}^L (s_j + 1)\cdot 3^{j-1}$, which maps each configuration to a unique integer in ${0,  \ldots, 3^L - 1}$. A general vector in the $L$-spin Hilbert space can be written as $|\psi_L\rangle = \sum_{n=0}^{3^L-1} c_n |n\rangle$, which is repeatedly acted upon by the transfer matrix in the eigenvalue solver: $\mathbf{T}|\psi_L\rangle$. The diagonal matrix $\mathbf{V}$ encodes the intra-row Boltzmann weights and is defined as
\begin{equation} \label{eq:V}
\langle \mathbf{s} | \mathbf{V} | \mathbf{s} \rangle =
e^{\beta J (s_L s_1 + \sum_{j=1}^{L-1} s_j s_{j+1}) - \beta\Delta\sum_{j=1}^L s_j^2},
\end{equation}
where the first term accounts for nearest-neighbor interactions with periodic boundary conditions, the second term incorporates the single-ion anisotropy, and $\beta \equiv 1/T$ is the inverse temperature. The symmetric form of $\mathbf{T}$ in Eq.~\eqref{eq:tmatrix} is achieved by inserting $\mathbf{V}^{1/2}$ on both sides of the transfer sequence $\mathbf{T}_L \cdots \mathbf{T}_1$.

Special care is required to correctly implement periodic boundary conditions in the first ($\mathbf{T}_1$) and last ($\mathbf{T}_L$) transfer operations, which correspond to the leftmost and rightmost sites of the row, respectively. For the first transfer at site $j=1$, the matrix element of $\mathbf{T}_1$ is given by
\begin{equation} \label{eq:T1}
    \langle \mathbf{s} | \mathbf{T}_1 | \mathbf{s^\prime} \rangle =
    e^{\beta J s_1 (s_1^\prime + s_2^\prime)} \,
    \delta_{s_{L+1}, s_1^\prime}
    \prod_{j=2}^L \delta_{s_j, s_j^\prime}\,,
\end{equation}
where the spin $s_1^\prime$ from the newly inserted row is stored in an auxiliary position $s_{L+1}$ to ensure correct interaction tracking. This effectively extends the spin configuration by one site, making the matrix $\mathbf{T}_1$ of size $3^{L+1} \times 3^L$. This step is illustrated in Fig.~\ref{fig:tm-method}(b), where the transition from the initial state $|s_{1,1} s_{1,2} \cdots s_{1,L} \rangle$ to the intermediate state $|s_{2,1} s_{1,2} \cdots s_{1,L} s_{1,1} \rangle$ is shown, accompanied by the Boltzmann weight encoding the inter-row interaction at $s_{2,1}$.

\begin{figure}
    \centering
    \includegraphics[width=0.85\linewidth]{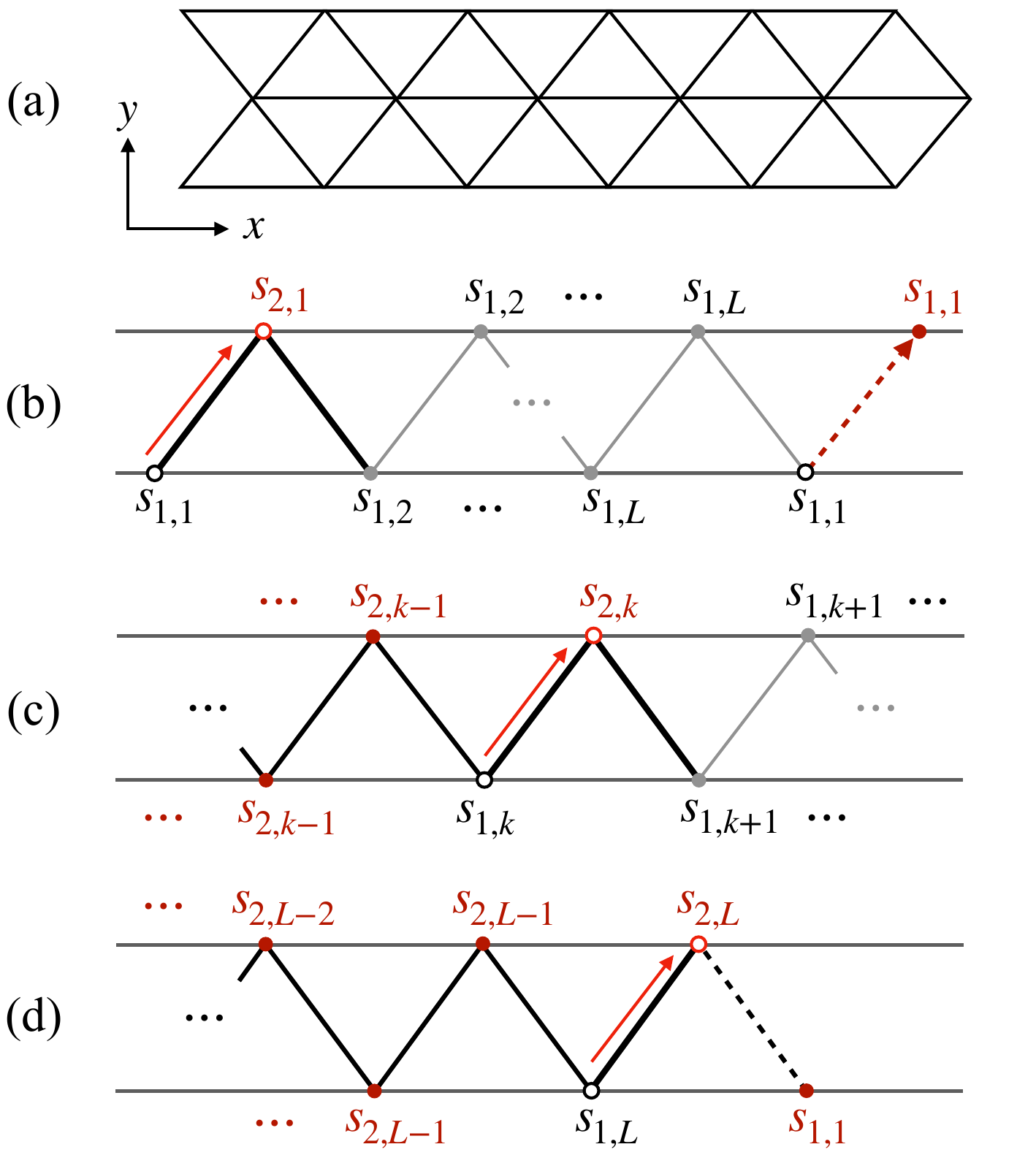}
    \caption{Lattice geometry and sparse matrix factorization. (a) Schematic of the symmetric transfer-matrix setup for an infinite strip oriented along the $y$-axis. The sparse matrix–vector multiplication begins with (b) $\mathbf{T}_1$, which transfers the leftmost spin and introduces an auxiliary spin. It proceeds through (c) $\mathbf{T}_k \cdots \mathbf{T}_1$, sequentially transferring the first $k$ spins, and concludes with (d) $\mathbf{T}_L \cdots \mathbf{T}_1$, completing the transfer of all spins across the row.}
    \label{fig:tm-method}
\end{figure}

The subsequent operations $\mathbf{T}_k \mathbf{T}_{k-1} \cdots \mathbf{T}_1$ incrementally build the intermediate state $|s_{2,1} s_{2,2} \cdots s_{2,k} s_{1,k+1} \cdots s_{1,L} s_{1,1}\rangle$ as illustrated in Fig.~\ref{fig:tm-method}(c). For $k = 2$ to $L-1$, the matrix element of $\mathbf{T}_k$ is given by
\begin{equation} \label{eq:Tk}
    \langle \mathbf{s} | \mathbf{T}_k | \mathbf{s^\prime} \rangle =
    e^{\beta J s_k (s_k^\prime + s_{k+1}^\prime)} \,
    \prod_{j = 1}^{k-1} \delta_{s_j, s_j^\prime}
    \prod_{j = k+1}^{L+1} \delta_{s_j, s_j^\prime}\,,
\end{equation}
where $s_k^\prime$ and $s_{k+1}^\prime$ represent newly added spins in the second row. The Kronecker deltas ensure that spins outside the update region remain unchanged. The matrix $\mathbf{T}_k$ acts on an extended spin configuration of size $L+1$, hence its dimension is $3^{L+1} \times 3^{L+1}$. The auxiliary spin $s_{L+1} = s^\prime_{L+1}$ is retained throughout the process to pass the stored value of $s_{1,1}$ to the final operator $\mathbf{T}_L$.

The auxiliary spin storing the initial state $s_{1,1}$ is finally utilized in the transfer of the rightmost spin site. The matrix element of $\mathbf{T}_L$
is given by
\begin{equation} \label{eq:TL}
    \langle \mathbf{s} | \mathbf{T}_L | \mathbf{s^\prime} \rangle =
    e^{\beta J s_L (s_L^\prime + s_{L+1}^\prime)} \,
    \prod_{j=1}^{L-1} \delta_{s_j, s_j^\prime}\,,
\end{equation}
where $s_{L+1}^\prime$ corresponds to the retrieved value of $s_{1,1}$. This operation completes the transformation to the full second-row configuration
$|s_{2,1} \cdots s_{2,L-1}, s_{2,L}\rangle$, as illustrated in Fig.~\ref{fig:tm-method}(d). The matrix $\mathbf{T}_L$ has dimensions $3^L \times 3^{L+1}$, reflecting that the auxiliary spin is discarded after use, thus restoring the original configuration size.

Since our geometry includes two layers of triangular plaquettes, the full transfer requires repeating the above operations to propagate spin configurations from the middle to the top row. This second set of transfers proceeds along the diagonal, as shown in Fig.~\ref{fig:tm-method}(a), which introduces a one-site shift to the right in the top row. To compensate for this shift and restore the intended geometry, we exploit the periodic boundary condition by applying a cyclic rotation operator $\mathbf{R}$, defined as
\begin{equation} \label{eq:R}
    \mathbf{R} | s_1 s_2 \cdots s_L \rangle = | s_L s_1 s_2 \cdots s_{L-1} \rangle\,.
\end{equation}
This rotation maps the state to the configuration depicted at the top row of Fig.~\ref{fig:tm-method}(a), thus ensuring that the resulting transfer matrix is symmetric. By incorporating this sparse-matrix factorization into the eigensolver~\cite{TRLan}, our matrix-free implementation takes full advantage of fast, on-the-fly sparse matrix–vector multiplications. These are parallelized using shared-memory computing and do not require explicit storage of the matrix elements.

We conclude this section by recalling that, in transfer-matrix analysis, phase transitions are signaled by the asymptotic degeneracy of the leading eigenvalues. It is important to emphasize that the structure of the transfer-matrix spectrum depends on the choice of boundary conditions, and that the number of quasi-degenerate eigenvalues reflects the nature of the low-temperature phase. In the present study of the triangular-lattice Blume–Capel model with periodic boundary conditions, we focus on the three lowest eigenvalues, as the tricritical point is characterized by the coexistence of three different states. In particular, the spin correlation length in the $y$ direction along the infinite strip is defined as
$\xi_{L} \equiv 1/\ln{(\lambda_1/\lambda_2)}$,
where $\lambda_1$ and $\lambda_2$ are the largest and second-largest eigenvalues, respectively. At a continuous phase transition, $\xi_L$ diverges due to the near degeneracy $\lambda_1 \approx \lambda_2$. In tricritical systems, the third-largest eigenvalue $\lambda_3$ also becomes important. At the tricritical point and along the first-order transition line, the largest three eigenvalues may become nearly degenerate~\cite{beale1986,xavier1998,rikvold1983,derrida1983,herrmann1984,beale1984}. To capture this, the so-called persistence length (also known as the second correlation length~\cite{derrida1983,herrmann1984}) is defined as
$\tilde{\xi}_{L} \equiv 1/\ln{(\lambda_1/\lambda_3)}$.
This quantity was introduced to detect the presence of metastable or coexisting phases~\cite{rikvold1983,derrida1983,herrmann1984,beale1984}. It is interpreted as a characteristic length scale of disordered domains along the infinite strip. The scaling behavior of $\tilde{\xi}_L$ is directly related to the interfacial tension between coexisting phases~\cite{rikvold1983}.

\begin{table}[ht]
  \centering
  \caption{Transition points along the phase boundary of the spin-$1$ Blume–Capel ferromagnet on the triangular lattice in the first-order transition regime, extending up to the tricritical point. The upper section of the table presents results from the transfer-matrix (TM) method obtained in the present work (second column), compared--where available--with previous estimates from two-parameter Wang–Landau (WL) (third column) and multicanonical (MUCA) (fourth column) simulations~\cite{mataragkas2025}. The lower section summarizes the most recent tricritical point estimates, emphasizing the superior numerical precision of the present transfer-matrix calculations. Numerical uncertainties in the TM results, of the order of $10^{-7}$, are omitted in the upper section for clarity and to avoid repetition.}
  \label{tab:pd}
  \medskip
    \begin{ruledtabular}
        \begin{tabular}{lccc} 
        \multicolumn{4}{c}{\textbf{Transitions Points}} \\ \toprule 
            $T$ &   & $\Delta^{\ast}$ & \\
           \hline
             & TM & WL & MUCA \\ \hline
            0.5000 & 2.998704 & 2.998691(8) & \\
            0.5500 & 2.997496 & 2.997482(6) &\\ 
            0.6000 & 2.995666 & 2.995600(5) & \\ 
            0.6500 & 2.992861 & 2.992852(4) & \\ 
            0.7000 & 2.989060 & 2.989055(4) & \\ 
            0.7500 & 2.984024 & 2.984020(4) & 2.98406(3) \\
            0.8000 & 2.977551 & 2.977568(3) & 2.97758(7) \\
            0.8500 & 2.969455 & 2.969520(2) & 2.96960(3) \\
            0.8750 & 2.964857 & &  \\
            0.9000 & 2.959711  & 2.959721(4) & 2.95972(1) \\
            0.9250 & 2.954137 & &  \\
            0.9500 & & 2.94808(1)  & 2.94795(3)\\
            0.9600 & & 2.94553(1)  &  \\
            0.9650 & & 2.94422(2)  & \\
            0.9700 & & 2.94290(2)  &  \\
            0.9750 & 2.941464 & 2.94157(2)  & \\
            0.9800 & & 2.94021(2)  & \\  
            0.9841 & 2.938909 &  &  \\
            0.9842 & 2.938881 &  &  \\
            0.9843 & 2.938853 &  &  \\
            0.9844 & 2.938825 &  &  \\
            0.9845 & 2.938796 & &  \\
            0.9846 & 2.938768 &  &  \\
            0.9847 & 2.938740 & &  \\ 
        \end{tabular}
    \end{ruledtabular}

  \begin{tabularx}{\linewidth}{X X X}
    \multicolumn{3}{c}{\textbf{Tricritical Point}} \\
    \toprule
    Ref. & $T_{\rm t}$ & $\Delta_{\rm t}$\\
    \midrule
    \cite{fytas_BC}  & 1.025(10) & 2.925(8) \\
    \cite{mataragkas2025}  & 0.9850(8) & 2.9388(2) \\
    Current work & 0.9847240(2) & 2.938568(1) \\
    \bottomrule
  \end{tabularx}
  
\end{table}

\section{Results}
\label{sec:results}

We start with a brief review of the finite-size scaling analysis of the correlation length, a widely used method for identifying critical points. The finite-size scaling ansatz for the correlation length can be expressed along the 
temperature axis as $\xi_L \approx L\, \mathcal{Q}(t L^{y_{\rm t}})$, where the scaling variable $t \equiv (T - T_{\rm c}) / T_{\rm c}$ quantifies the reduced temperature, $y_{\rm t} = 1/\nu$ is the thermal renormalization-group exponent, and $\mathcal{Q}$ is a universal scaling function~\cite{beale1986}. Similar expressions can be written for the case of analysis along the crystal-field axis as $\xi_L \approx L\, \mathcal{P}(\delta L^{y_{\rm t}})$,
where the scaling variable $\delta \equiv (\Delta - \Delta_{\rm c}) / \Delta_{\rm c}$ quantifies now the reduced crystal field, and $\mathcal{P}$ is another universal scaling function.  One can thus probe the system along the temperature axis $T$ at a fixed crystal field $\Delta$ (or along the crystal-field axis $\Delta$ at a fixed temperature $T$), estimating the critical value $T_{\rm c}$ (or $\Delta_{\rm c})$ by identifying the universal crossing point of the curves of $\xi_{L} / L$ plotted for different strip widths $L$, provided that $L$ is sufficiently large. In practice, however, the crossing points obtained from finite systems--especially those with relatively small $L$ values typical of transfer-matrix studies--do not coincide exactly due to finite-size corrections. This discrepancy is commonly addressed by performing a power-law extrapolation of the crossing temperatures, allowing for an accurate estimation of the critical point as well as the universal value of the ratio $\xi_{L}/L$ in the thermodynamic limit~\cite{derrida1982,luck1985,herrmann1984,blote1982,blote1988,privman1983,queiroz2000,night1996,ballesteros1996}. 

We determined the crossing point between two curves corresponding to strip widths $L$ and $L'$ by numerically finding a root in $T$ for a given value of $\Delta$ in the equation
\begin{equation}
\label{eq:quotients}
\frac{\xi_{L}(T,\Delta)}{L} = \frac{\xi_{L'}(T,\Delta)}{L'},
\end{equation}
where $L'$ is typically set to $L + 1$ or $L + 2$ throughout this work. The resulting crossing temperatures $T_{L}^{\ast}$ can be extrapolated using the well-known finite-size scaling relation~\cite{amit_book}
\begin{equation}
\label{eq:fss_T}
T_{L}^{\ast} = T_{\rm c} + bL^{-\tilde{\omega}},
\end{equation}
where $\tilde{\omega} \equiv \omega + y_{\rm t}$ combines the leading irrelevant scaling exponent $\omega$ and the shift exponent $y_{\rm t}$~\cite{derrida1982,privman1983,luck1985}. A similar analysis applies to the crossing points in the crystal field, $\Delta_{L}^{\ast}$, which follow an analogous scaling form to Eq.~\eqref{eq:fss_T}.

\begin{figure}
    \centering
    \includegraphics[width=1.0\linewidth]{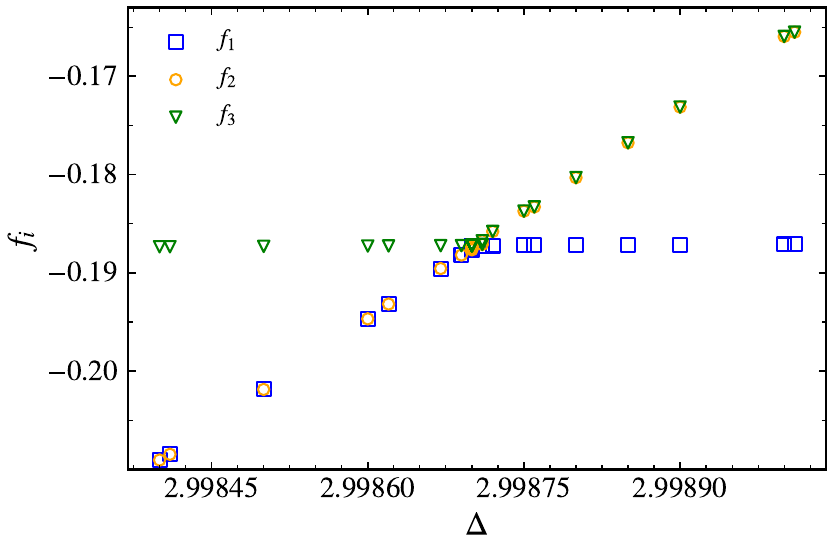}
    \caption{The lowest three free-energy levels corresponding to $\lambda_{1}$ > $\lambda_{2}$ > $\lambda_{3}$, plotted at $T = 0.5$ for $L = 18$ around the determined transition point $\Delta^{\ast} = 2.99870374$. Notably, for $\Delta < \Delta^{\ast}$, the levels $f_1$ and $f_2$ are nearly degenerate, reflecting the standard ferromagnetic phase. In contrast, for $\Delta > \Delta^{\ast}$, the hierarchy changes to $f_1 < f_2 \approx f_3$, indicating a disordered, vacancy-dominated phase.}    \label{fig:free_energy}
\end{figure}

It is worth noting that the finite-size scaling analysis of the correlation length given above implicitly assumes the presence of a second-order phase transition. One may strictly say that such a scaling ansatz is not applicable to first-order transitions. However, it has been witnessed that the same extrapolation strategy with crossing points between the curves of adjacent system sizes still provides a reasonable estimate of transition in the vicinity of the tricritical point~\cite{beale1986,rikvold1983}.

\subsection{Free-energy level spectrum}
\label{sec:free-energy}

The free-energy levels are defined as $f_{i} \equiv -\frac{T}{(\sqrt{3}L)} \ln{\lambda_{i}}$, where $\lambda_i$ denotes the i\textsuperscript{th} largest eigenvalue of the transfer matrix and the geometrical prefactor $\sqrt{3}$ accounts for the strip orientation and the three-row double-layer construction shown in Fig.~\ref{fig:tm-method}(a). These levels provide a powerful diagnostic tool for examining the behavior of equilibrium and metastable phases across a first-order phase transition. In tricritical Ising-like systems~\cite{beale1986,xavier1998,rikvold1983,bartelt1986}, the coexistence of two ordered and one disordered phase manifests as an abrupt rearrangement in the leading eigenvalues, resulting in distinct features in the corresponding free-energy levels. Such spectral signatures have been previously observed in two-dimensional lattice gas models~\cite{rikvold1983,bartelt1986}, where they reveal the underlying phase coexistence. More recently, the expected exponential system-size scaling of spectral degeneracy--a hallmark of first-order transitions--was explicitly demonstrated in the square-lattice ferromagnetic Blume-Capel model~\cite{kim17}. Here, we extend this understanding by providing direct evidence for this scaling behavior in the triangular-lattice version of the model.

Figure~\ref{fig:free_energy} shows the behavior of the three lowest free-energy levels at $T = 0.5$ as a function of the crystal field $\Delta$ across the first-order transition regime. A pronounced avoided crossing in the equilibrium free energy, corresponding to the lowest level, is observed at $\Delta = \Delta^{\ast} = 2.99870374$, signaling a first-order phase transition. For $\Delta > \Delta^{\ast}$, the system resides in the disordered phase, characterized by a non-degenerate ground state well separated from the next two excited levels. In contrast, for $\Delta < \Delta^{\ast}$, the two lowest levels become nearly degenerate, indicating the coexistence of two ordered phases. As the strip width $L$ increases, this change in the free-energy level structure
at $\Delta = \Delta^{\ast}$ becomes increasingly sharp, enabling precise determination of the transition point.

Although the free-energy level crossing appears evident at $L = 18$, the gap between levels remains finite for any finite system size. This avoided crossing can be quantitatively characterized by examining the system-size scaling of the difference between $f_1$ and $f_3$, as shown in Fig.~\ref{fig:spectral_gap}. We find that this gap at $\Delta^{\ast}$ decreases exponentially with increasing $L$, indicating that while the levels do not become exactly degenerate at finite $L$, a kink in the free-energy landscape emerges and sharpens exponentially with system size. Identifying the value of $\Delta^{\ast}$ at which this gap is minimized provides a reliable estimate of the first-order transition point. This exponential closing of the spectral gap is directly related to the divergence of the correlation and persistence lengths, $\xi$ and $\tilde{\xi}$, respectively.  

\begin{figure}
    \centering
    \includegraphics[width=1.0\linewidth]{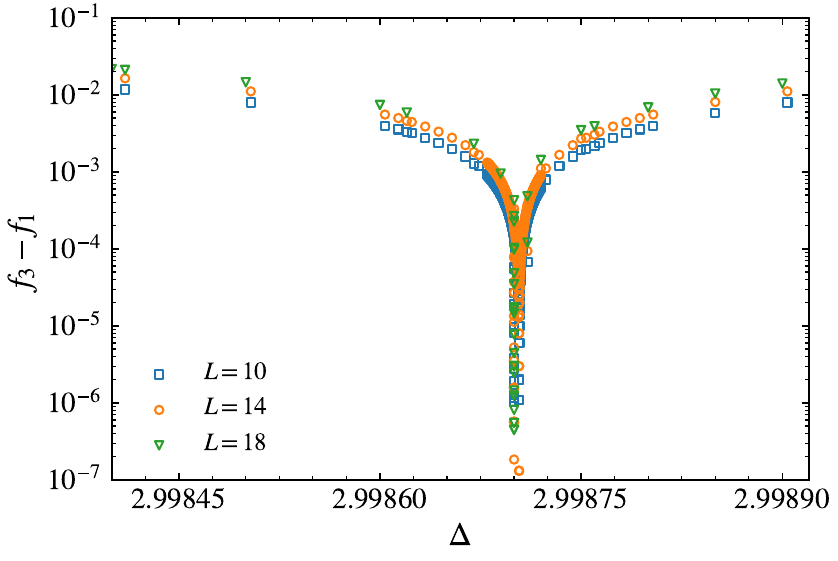}
    \caption{Spectral gap illustration of the difference between $f_{1}$ and $f_{3}$ at $T = 0.5$ with increasing system size.}    \label{fig:spectral_gap}
\end{figure}

At very low temperatures,
the location of $\Delta^{\ast}$--identified via the minimum gap--turns out to be almost independent of
the strip width $L$, as exemplified
in the case of $T = 0.5$ in Fig.~\ref{fig:spectral_gap}. For $T \leq 0.5$, the finite-size correction is very much suppressed in $\Delta^\ast$, varying
by less than $10^{-8}$ across the range of system sizes considered. As the temperature increases towards the tricritical point, the finite-size correction becomes visible. However, a simple power-law extrapolation provides a reliable estimate of a transition point in the entire first-order area. Compared to the Wang–Landau method, which demands careful treatment of the field-mixing analysis, this minimum-gap approach offers a straightforward and highly reliable alternative to construct the phase coexistence curve.
Notably, for the specific case of $T = 0.5$ considered here, the value $\Delta^{\ast} = 2.998691(8)$ obtained from Wang–Landau simulations with refined field mixing~\cite{mataragkas2025} is in excellent agreement with the value $\Delta^{\ast} = 2.99870374$ determined in this work.

\subsection{Interfacial tension and the tricritical point}
\label{sec:free-energy}

\begin{figure}
    \centering
    \includegraphics[width=1.0\linewidth]{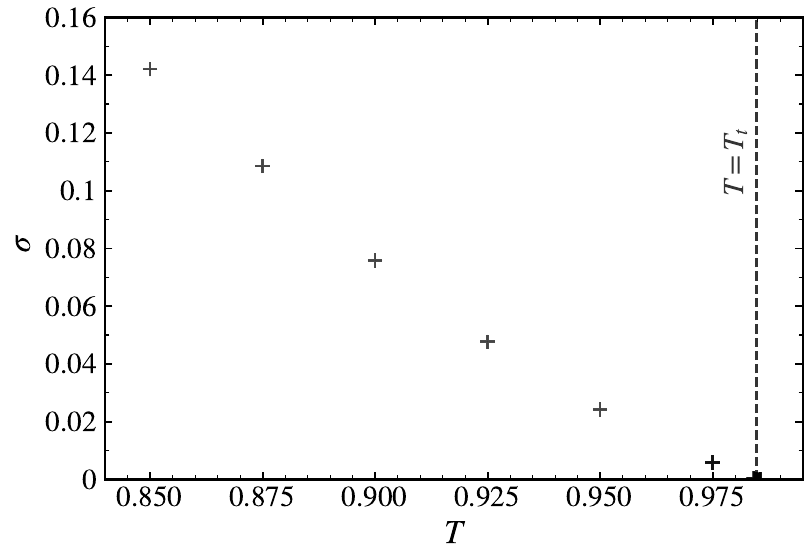}
    \caption{Interfacial tension $\sigma$ as a function of temperature. The value decreases to below $2.5 \times 10^{-6}$ at $T = 0.9847$, approaching the tricritical temperature $T_{\rm t}$, indicated by the vertical dashed line.}    \label{fig:interface_tension}
\end{figure}

In our free-energy level calculations, for $\Delta < \Delta^{\ast}$, the difference $(f_1 - f_2)$ decays exponentially with increasing strip width $L$, while $(f_1 - f_3)$ saturates to a finite value. In contrast, for $\Delta > \Delta^{\ast}$, the gap $(f_2 - f_3)$ decays exponentially, whereas $(f_1 - f_3)$ remains finite. Exactly at the transition point $\Delta = \Delta^{\ast}$, all three level differences decay exponentially with the system size. This simultaneous exponential behavior is used to locate the first-order transition points. On the other hand, exactly at the tricritical point these level differences converge to a fixed value that can be predicted from conformal field theory, a strategy we also adopt in the present work. When rephrased in terms of the previously introduced length scale, this spectral behavior directly reflects the exponential scaling of the persistence length $\tilde{\xi}^\ast$ with $L$. It is well established that the asymptotic growth of the persistence length is closely related to the interfacial tension between coexisting phases at a first-order transition (see Refs.~\cite{rikvold1983,privman1983} and references therein). The relationship is formally given by
\begin{equation}
\label{eq:interface_tension}
\tilde{\xi}^{\ast}_L \sim L \exp{[\beta \sigma L]},
\end{equation}
where $\sigma$ denotes the interfacial tension.

Along the phase coexistence curve--identified either by the peak value of $\tilde{\xi}^{\ast}_{L}$ or equivalently by the minimum gap between the free-energy levels--we extract the temperature dependence of the interfacial tension, as shown in Fig.~\ref{fig:interface_tension}. The values of $\sigma$ are deduced from the exponential growth behavior of $\tilde{\xi}^{\ast}_{L}$, exemplified in Fig.~\ref{fig:fss}. An interesting feature of the interfacial tension is its almost linear increase with the temperature as the system moves away from the tricritical point into the first-order regime. This trend is consistent with previous transfer-matrix studies of the square-lattice Blume-Capel model~\cite{kim17}, as well as with multicanonical simulations of the triangular-lattice model~\cite{mataragkas2025}. As the interfacial tension $\sigma$ grows, the corresponding length scale diverges more rapidly with increasing strip width $L$. This accelerated divergence not only accounts for the precise determination of the transition field $\Delta^{\ast}$ using relatively modest strip widths, but also implies that signatures of discontinuity in thermodynamic quantities become more evident even in small systems. In particular, the enhanced rate of divergence leads to a sharper development of a kink in the free energy, providing a clear indication of first-order behavior.

\begin{figure}
    \centering
    \includegraphics[width=1.0\linewidth]{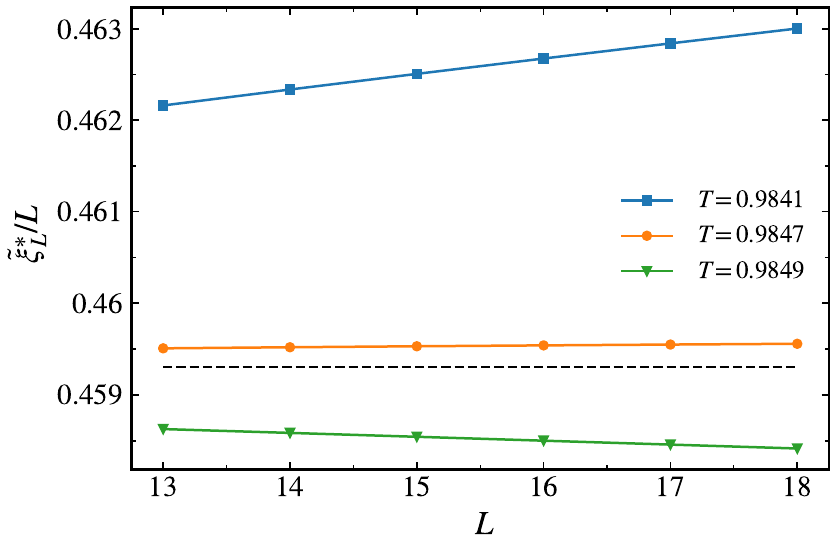}
    \caption{Exponential scaling behavior of the scaled persistence length, $\tilde{\xi}^{\ast}_{L}/L$, across the first-order and second-order transition regimes, with the tricritical temperature $T_{\rm t}$ indicated by the dashed horizontal line.}  \label{fig:fss}
\end{figure}

This trend is corroborated by a finite-size scaling analysis around the tricritical region, see Fig.~\ref{fig:tricritical_point_fss}(a), where the rescaled ratio $\tilde{\xi}^{\ast}_{L}/L$ is plotted as a function of the temperature for the whole range of strip widths considered. In Fig.~\ref{fig:tricritical_point_fss}(b) we present the extrapolation of the crossing temperatures using the scaling form~\eqref{eq:fss_T}. A best fit to the data, fixing $\tilde{\omega} = 3$, yields $T_{\rm t} = 0.9847240(2)$. A similar analysis along the crystal-field coupling direction, see Fig.~\ref{fig:tricritical_point_fss}(c), yields $\Delta_{\rm t} = 2.938568(1)$, where in this case $\tilde{\omega} \approx 2.8$ (in particular $\tilde{\omega} = 2.847$ for $N = L$ and $2.793$ for $N = L+1$). Note that since $y_{\rm t} = 1/\nu = 9/5$, corresponding to the Ising tricritical universality class~\cite{kwak2015,kim17, mataragkas2025}, the above fit results for $\tilde{\omega}$ suggest an empirical value of $\omega \approx 1.0$. We thus arrive at a final estimate for the tricritical point: $(\Delta_{\rm t}, T_{\rm t}) = [2.938568(1), 0.9847240(2)]$. This improves by several order of magnitude in accuracy the most advanced Monte Carlo estimate $(\Delta_{\rm t}, T_{\rm t}) = [2.9388(2), 0.9850(8)]$~\cite{mataragkas2025}.

\begin{figure}
    \centering
    \includegraphics[width=1.0\linewidth]{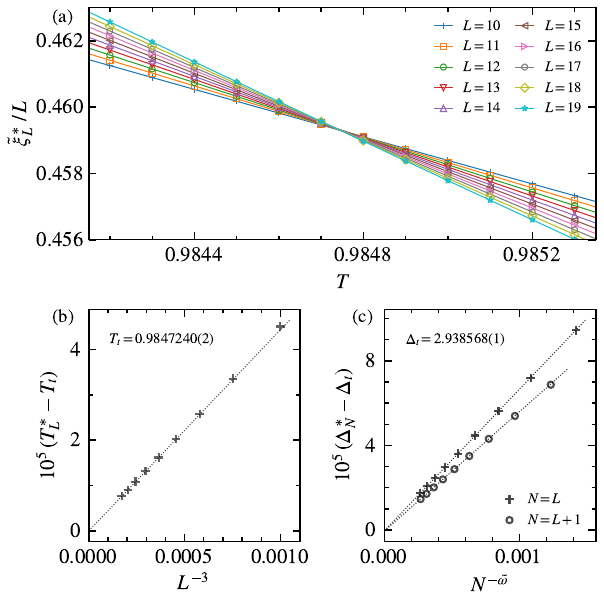}
    \caption{Finite-size scaling analysis for determining the tricritical point.
(a) The maximum persistence length $\tilde{\xi}_L^\ast$ is shown as a function of temperature $T$ for various system sizes $L$.
(b) The crossing temperatures $T^{\ast}_L$ between successive pairs of system sizes $(L, L+1)$ are extrapolated to estimate the tricritical temperature $T_{\rm t}$ in the thermodynamic limit.
(c) The corresponding pseudo-transition fields $\Delta^{\ast}_L(T^{\ast}_L)$ and $\Delta^{\ast}_{L+1}(T^{\ast}_{L+1})$ are used to extract the tricritical field $\Delta_{\rm t}$ via a power-law extrapolation.}
    \label{fig:tricritical_point_fss}
\end{figure}

\subsection{Conformal field theory validation}
\label{sec:universal_ratio}

We demonstrate the accuracy of our tricritical point estimate by confirming its consistency with the predictions of conformal field theory for the tricritical Ising universality class~\cite{henkel_book}. Remarkably, such theoretical validations were first carried out nearly four decades ago by Alcaraz \emph{et al.}~\cite{alcaraz85}, within the time-continuum formulation of the model. While it is well established that the tricritical point of the two-dimensional spin-$1$ Blume–Capel model belongs to this universality class, most previous numerical studies have focused on the square-lattice version to test conformal field theory predictions and to aid in locating the tricritical point~\cite{xavier1998, qian05, deng05, mandal16, belov17, blote19}. In what follows, we present conformal field theory consistency checks for the triangular-lattice Blume–Capel model. Specifically, we provide a precise verification of the central charge and scaling dimensions using the transfer-matrix spectrum evaluated at our estimate of the tricritical point.

The central charge $c$ can be extracted from the finite-size scaling behavior of the free energy. For a cylindrical geometry--specifically, an infinitely long strip with periodic boundary conditions in the transverse direction--the free-energy density $f$, expressed in units of $k_{\rm B}T$, approaches the bulk value $f_\infty$ as the strip width $L$ increases, following the well-known scaling relation
\begin{equation} \label{eq:fss_free_energy}
    f(L) = f_\infty - \frac{\pi c}{6L^2} + O(L^{-4})\,,
\end{equation}
where $c$ is the central charge characterizing the underlying conformal field theory. By tracking this finite-size dependence numerically, one can accurately determine $c$ from the computed values of $f(L)$. In our implementation for the triangular-lattice Blume–Capel model (see Fig.~\ref{fig:tm-method}(a) and the relevant discussion in Sec.~\ref{sec:free-energy}), the free-energy density is obtained from the largest eigenvalue $\lambda_1$ of the transfer matrix as
\begin{equation}
    f(L) = -\frac{1}{\sqrt{3}L} \ln \lambda_1\,.
\end{equation}

To extract the central charge, we first estimate the bulk free energy density $f_\infty$ using the Bulirsch-Stoer extrapolation algorithm~\cite{henkel88} with the control parameter $\omega = 2$. This enables us to construct the finite-size estimator
\begin{equation}
    c(L) = \frac{6L^2}{\pi}\left[f_\infty - f(L)\right]\,,
\end{equation}
from which the central charge is obtained by extrapolating $c(L)$ to the thermodynamic limit $L \to \infty$. Figure~\ref{fig:CFT}(a) shows a linear fit to the form $c(L) = c + a L^{-1}$ for system sizes $L > 12$, yielding an estimate of $c = 0.7000167(2)$. The uncertainty in parentheses denotes the standard deviation obtained from jackknife resampling, where each fit omits a single data point to assess the robustness of the extrapolation. For comparison, applying the Bulirsch-Stoer extrapolation with $\omega = 1$ gives a closely matching estimate of $c = 0.700000(1)$, where the error is defined as twice the difference between the last two approximants in the sequence. Both estimates are in excellent agreement with the exact value $c = 7/10$ predicted by conformal field theory for the tricritical Ising universality class.

\begin{figure}
    \centering
    \includegraphics[width=1.0\linewidth]{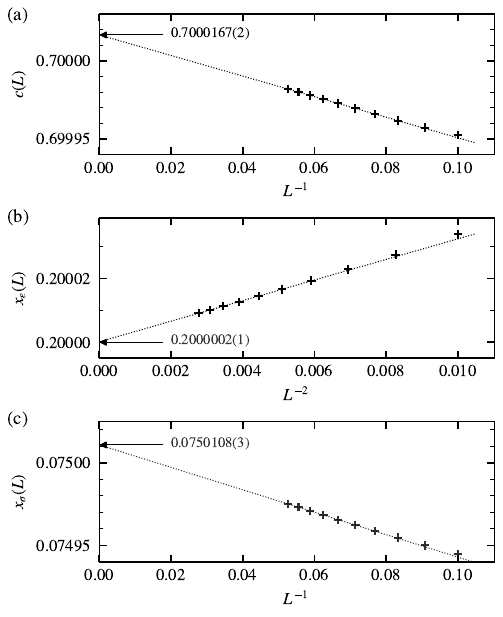} 
\caption{Conformal field theory verification of the tricritical Ising universality class.
(a) The central charge $c$, (b) the thermal scaling dimension $x_\epsilon$, and (c) the magnetic scaling dimension $x_\sigma$ are extracted from the transfer-matrix spectrum at $T = 0.9847240$ and $\Delta = 2.938568$ for strip widths up to $L = 19$. Linear extrapolations (arrows) are compared against the exact conformal field theory values: $c = 7/10$, $x_\epsilon = 1/5$, and $x_\sigma = 3/40$.}
    \label{fig:CFT}
\end{figure}

The thermal and magnetic scaling dimensions, $x_\epsilon$ and $x_\sigma$, are related to the universal ratios involving the persistence length $\tilde{\xi}_L$ and the correlation length $\xi_L$, respectively, through the finite-size scaling relations
\begin{equation}
    2\pi x_\epsilon = \frac{L}{\sqrt{3}\, \tilde{\xi}_L}
    \quad \text{and} \quad
    2\pi x_\sigma = \frac{L}{\sqrt{3}\, \xi_L}
\end{equation}
in the limit of large system size $L$. As already noted above, the geometric factor $\sqrt{3}$ accounts for the effective unit length along the cylindrical axis, determined by our choice of a double-layered structure in the transfer-matrix construction.

The estimates of the scaling dimensions obtained from the extrapolations shown in Figs.~\ref{fig:CFT}(b) and (c) show excellent agreement with the exact values $x_\epsilon = 1/5$ and $x_\sigma = 3/40$. For the thermal scaling dimension $x_\epsilon$, the data points for $L > 12$ exhibit a nearly perfect $L^{-2}$ behavior, yielding an extrapolated value of $x_\epsilon = 0.2000002(1)$ in the thermodynamic limit. This result is consistent with the Bulirsch-Stoer extrapolation using control parameter $\omega = 2$, which provides $x_\epsilon = 0.200002(2)$. In the case of the magnetic scaling dimension $x_\sigma$, the finite-size behavior appears to follow an approximate $L^{-1}$ form. A linear extrapolation using data for $L > 12$ yields $x_\sigma = 0.0750108(3)$. The Bulirsch-Stoer extrapolation with $\omega = 1$ results in  $x_\sigma = 0.0749(1)$, where the slightly larger deviation suggests the presence of stronger finite-size corrections. To further verify the robustness of our estimate, we also performed a power-law fit without assuming a specific correction form, which produced $x_\sigma = 0.075035(1)$, demonstrating consistency across different extrapolation methods.
The excellent agreement with the conformal field theory predictions--namely, $c = 7/10$, $x_\epsilon = 1/5$, and $x_\sigma = 3/40$--provides strong support for the accuracy of our tricritical point estimate, given by $(\Delta_{\rm t}, T_{\rm t}) = [2.938568(1), 0.9847240(2)]$.

\section{Summary and Outlook}
\label{sec:conclusions}

In this work, we conducted a high-precision transfer-matrix analysis of the spin-$1$ Blume–Capel model on the triangular lattice. Implementing the sparse-matrix factorization for a symmetric strip geometry with periodic boundary conditions,
we extended our computations to strip widths up to $L = 19$. This enabled a detailed mapping of the phase diagram--a summary of transition points
is provided in Table~\ref{tab:pd}--with particular emphasis on the accurate determination of the tricritical point at $(\Delta_{\rm t}, T_{\rm t}) = [2.938568(1), 0.9847240(2)]$. We confirmed the accuracy
of the tricritical point by validating
the central charge and scaling dimensions with
the transfer-matrix spectrum, achieving excellent agreement with conformal field theory predictions for the tricritical Ising class. In the first-order regime, we analyzed the scaling behavior of the spectral gap to determine the interfacial tension, revealing its linear increase as temperature decreases--consistent with theoretical expectations and prior results. Our construction of the phase coexistence curve and rigorous finite-size scaling analysis further bring out the power of the transfer-matrix method to resolve fine thermodynamic structure. Notably, our tricritical point estimate aligns remarkably well with the most precise Monte Carlo results to date, providing a compelling cross-validation between distinct numerical methodologies.

Altogether, this study reinforces the transfer-matrix method as a robust and versatile tool for probing both continuous and first-order phase transitions, especially in models with complex geometries. By integrating 
sparse-matrix techniques with meticulous scaling analysis, we deliver results of benchmark-level accuracy that advance our understanding of critical phenomena. Looking forward, we are extending this framework to more intricate systems, such as models with non-reciprocal interactions introduced through asymmetric
coupling matrices, relevant to active-matter-inspired dynamics, where conventional simulation methods often struggle. These ongoing efforts aim to expand the applicability of transfer-matrix approaches to a broader class of challenging problems in statistical and condensed matter physics.

\begin{acknowledgments}
We are grateful to Professor Per Arne Rikvold for the very useful correspondence regarding our work. Part of the numerical calculations reported in this paper were performed at the High-Performance Computing cluster CERES of the University of Essex. The work of A.V. and N.G.F. was supported by the  Engineering and Physical Sciences Research Council (grant EP/X026116/1 is acknowledged). D.H.K. acknowledges the support from the National Research Foundation of Korea grant (No. RS-2024-00392445) funded by the Korea government (MSIT).
\end{acknowledgments}

\bibliography{biblio}

\end{document}